\documentclass[preprintnumbers,amsmath,amssymbm,prd]{revtex4}
\usepackage{epsfig}
\usepackage{graphicx}

\begin{document}
\title{The charged black-hole bomb: A lower bound on the charge-to-mass ratio of the explosive scalar field}
\author{Shahar Hod}
\affiliation{The Ruppin Academic Center, Emeq Hefer 40250, Israel}
\affiliation{ } \affiliation{The Hadassah Institute, Jerusalem
91010, Israel}
\date{\today}

\begin{abstract}
\ \ \ The well-known superradiant amplification mechanism allows a
charged scalar field of proper mass $\mu$ and electric charge $q$ to
extract the Coulomb energy of a charged Reissner-Nordstr\"om black
hole. The rate of energy extraction can grow exponentially in time
if the system is placed inside a reflecting cavity which prevents
the charged scalar field from escaping to infinity. This composed
black-hole-charged-scalar-field-mirror system is known as the {\it
charged black-hole bomb}. Previous numerical studies of this
composed physical system have shown that, in the linearized regime,
the inequality $q/\mu>1$ provides a necessary condition for the
development of the superradiant instability. In the present paper we
use analytical techniques to study the instability properties of the
charged black-hole bomb in the regime of linearized scalar fields.
In particular, we prove that the lower bound
${{q}\over{\mu}}>\sqrt{{{r_{\text{m}}/r_--1}\over{r_{\text{m}}/r_+-1}}}$
provides a necessary condition for the development of the
superradiant instability in this composed physical system (here
$r_{\pm}$ are the horizon radii of the charged Reissner-Nordstr\"om
black hole and $r_{\text{m}}$ is the radius of the confining
mirror). This {\it analytically} derived lower bound on the
superradiant instability regime of the composed
black-hole-charged-scalar-field-mirror system is shown to agree with
direct {\it numerical} computations of the instability spectrum.
\end{abstract}
\bigskip
\maketitle


\section{Introduction}

Kerr black holes may contain large amounts of rotational energies
which can be released by bosonic fields that scatter off these
spinning black holes. In this physical process, which is known as
superradiant scattering \cite{Zel,PressTeu2,Viln}, an incident
bosonic field whose proper frequency lies in the superradiant regime
\cite{Zel,PressTeu2,Viln,Noteun}
\begin{equation}\label{Eq1}
0<\omega<m\Omega_{\text{H}}\
\end{equation}
can extract the rotational energy and angular momentum of the
spinning Kerr black hole (Here $m$ is the azimuthal harmonic index
of the incident bosonic field and $\Omega_{\text{H}}$ is the angular
velocity of the black-hole horizon).

The rate of energy extraction from the black hole can grow
exponentially in time if the bosonic field is prevented from
escaping to infinity. The required confinement mechanism can be
provided either by a reflecting mirror which surrounds the black
hole \cite{PressTeu2,CarDias} or, for a massive bosonic field, by
the mutual gravitational attraction between the central black hole
and the extracting field \cite{Notemas,HerR}.

It should be emphasized that not all bosonic modes trigger the
black-hole superradiant instability. In particular, it was proved in
\cite{Hodbound} that the inequality
\begin{equation}\label{Eq2}
\mu<\sqrt{2}\cdot m\Omega_{\text{H}}\
\end{equation}
provides a necessary condition for the development of the
superradiant instability in the composed
Kerr-black-hole-massive-scalar-field system, where $\mu$ is the
proper mass of the exploding scalar field.

As pointed out by Bekenstein \cite{Bekch}, an analogous superradiant
amplification of bosonic fields may occur when a {\it charged} field
scatters off a {\it charged} black hole. In particular, a charged
scalar field whose proper frequency lies in the superradiant regime
\cite{Bekch}
\begin{equation}\label{Eq3}
0<\omega<q\Phi_{\text{H}}\
\end{equation}
can extract the Coulomb energy and electric charge of a charged
Reissner-Nordstr\"om (RN) black hole (Here $q$ is the charge
coupling constant of the incident scalar field and $\Phi_{\text{H}}$
is the electric potential of the charged black hole).

Interestingly, it was proved in \cite{Hodnbt} that, contrary to the
spinning (Kerr) case, in the charged (RN) case the gravitational
attraction between the black hole and the massive charged scalar
field {\it cannot} provide the confinement mechanism which is
required in order to trigger the black-hole superradiant
instability. The charged black-hole bomb must therefore include a
reflecting mirror which surrounds the black hole and prevents the
amplified charged bosonic field from escaping to infinity
\cite{Dego,Hodexp,Lim}.

In a very interesting work, Degollado et. al. \cite{Dego} have used
numerical techniques to study the instability properties of the
composed RN-black-hole-charged-scalar-field-mirror system. In
particular, it was found in \cite{Dego} that, in the linearized
regime \cite{Notelin}, the inequality
\begin{equation}\label{Eq4}
{{q}\over{\mu}}>1\
\end{equation}
provides a necessary condition for the development of the
superradiant instabilities in this charged system.

The main goal of the present paper is to explore the superradiant
instability regime of the composed
RN-linearized-charged-scalar-field-mirror system (the charged
black-hole bomb) using analytical techniques. In particular, below
we shall provide an {\it analytical} explanation for the
characteristic inequality (\ref{Eq4}) observed {\it numerically} in
the interesting study of Degollado et. al. \cite{Dego}. Moreover, in
this paper we shall derive a stronger lower bound [see Eq.
(\ref{Eq45}) below] on the dimensionless charge-to-mass ratio which
characterizes the explosive charged massive scalar fields.

\section{Description of the system}

We shall study the dynamics of a charged massive scalar field $\Psi$
linearly coupled to a non-extremal charged RN black hole of mass $M$
and electric charge $Q$. The charged RN black-hole spacetime is
described by the line element \cite{Chan,Noterr}
\begin{equation}\label{Eq5}
ds^2=-f(r)dt^2+{1\over{f(r)}}dr^2+r^2(d\theta^2+\sin^2\theta
d\phi^2)\ ,
\end{equation}
where the metric function $f(r)$ is given by
\begin{equation}\label{Eq6}
f(r)= 1-{{2M}\over{r}}+{{Q^2}\over{r^2}}\  .
\end{equation}
The zeros of $f(r)$,
\begin{equation}\label{Eq7}
r_{\pm}=M\pm (M^2-Q^2)^{1/2}\  ,
\end{equation}
determine the horizon radii of the charged RN black hole
\cite{Chan}.

The dynamics of a test scalar field $\Psi$ of proper mass $\mu$ and
charge coupling constant $q$ \cite{Noteuni,Noteqp} in the background
of the RN black-hole spacetime is governed by the Klein-Gordon wave
equation \cite{HodPirpam,Stro,HodCQG2}
\begin{equation}\label{Eq8}
[(\nabla^\nu-iqA^\nu)(\nabla_{\nu}-iqA_{\nu}) -\mu^2]\Psi=0\  ,
\end{equation}
where $A_{\nu}=-\delta_{\nu}^{0}{Q/r}$ is the electromagnetic
potential of the charged black hole. One can decouple the radial and
angular parts of the scalar field $\Psi$ and express it in the form
\begin{equation}\label{Eq9}
\Psi_{lm}(t,r,\theta,\phi)=e^{im\phi}S_{lm}(\theta)R_{lm}(r)e^{-i\omega
t}\ ,
\end{equation}
where $\omega, l$, and $m$ are respectively the conserved frequency
of the field mode and its angular harmonic indices
\cite{Notelm,Noteom}.

It is worth noting that the sign of $\Im\omega$ in (\ref{Eq9})
determines the (in)stability properties of the scalar field mode:
stable modes (that is, modes decaying in time) are characterized by
$\Im\omega<0$, whereas unstable modes (that is, modes growing
exponentially in time) are characterized by $\Im\omega>0$.
Stationary modes with $\Im\omega=0$ mark the boundary between stable
and unstable solutions of the Klein-Gordon wave equation
(\ref{Eq8}). These marginally stable field modes are characterized
by the critical (marginal) frequency [see Eq. (\ref{Eq3})]
\begin{equation}\label{Eq10}
\omega_{\text{c}}={{qQ}\over{r_+}}\
\end{equation}
for the superradiant scattering phenomenon \cite{Bekch}.

Substituting the scalar field decomposition (\ref{Eq9}) into the
Klein-Gordon wave equation (\ref{Eq8}) and using the line element
(\ref{Eq5}) of the RN black-hole spacetime, one finds that the
radial function $R(r)$ is determined by the characteristic radial
equation \cite{HodPirpam,Stro,HodCQG2}
\begin{equation}\label{Eq11}
\Delta{{d} \over{dr}}\Big(\Delta{{dR}\over{dr}}\Big)+UR=0\ ,
\end{equation}
where
\begin{equation}\label{Eq12}
\Delta\equiv r^2f(r)\
\end{equation}
and
\begin{equation}\label{Eq13}
U\equiv(\omega r^2-qQr)^2 -\Delta(\mu^2r^2+K_l)\  .
\end{equation}
Here $K_l=l(l+1)$ (where $m$ and $l\geq |m|$ are integers) are the
characteristic eigenvalues of the angular function $S(\theta)$
\cite{HodPirpam,Stro,HodCQG2}.

The characteristic equation (\ref{Eq11}) for the radial
eigenfunction $R(r)$ should be supplemented by the physical boundary
condition of purely ingoing waves at the black-hole horizon
\cite{CarDias,Notemas,HerR,Dego}:
\begin{equation}\label{Eq14}
R \sim e^{-i (\omega-qQ/r_+)y}\ \ \text{ as }\ r\rightarrow r_+\ \ \
(y\to -\infty)\  ,
\end{equation}
where the radial coordinate $y$ is determined by the relation
$dy=dr/f(r)$ [see Eq. (\ref{Eq18}) below]. For field modes in the
superradiant regime (\ref{Eq3}), the near-horizon boundary condition
(\ref{Eq14}) corresponds to an outgoing flux of Coulomb energy and
electric charge from the charged RN black hole
\cite{Bekch,Hodnbt,Dego,Hodexp,Lim}. In addition, the reflecting
mirror which surrounds the composed black-hole-field system dictates
the boundary condition \cite{CarDias,Dego,Hodexp,Lim}
\begin{equation}\label{Eq15}
R(r=r_{\text{m}})=0\
\end{equation}
for the confined scalar field, where $r_{\text{m}}$ is the radial
location of the mirror.

\section{The effective radial potential of the composed RN-black-hole-charged-massive-scalar-field system}

The radial equation (\ref{Eq11}), together with the boundary
conditions (\ref{Eq14}) and (\ref{Eq15}), determine a discrete
family of complex field resonances $\{\omega_n(r_{\text{m}})\}$
\cite{CarDias,Dego,Hodexp,Noteqnm}. As mentioned above, Degollado
et. al. \cite{Dego} have performed a very interesting numerical
study of these characteristic resonances of the composed
RN-black-hole-charged-scalar-field-mirror system. In particular,
Degollado et. al. \cite{Dego} have found numerically that unstable
(exploding) charged field modes are characterized by the property
$q>\mu$ [see Eq. (\ref{Eq4})]. The main goal of the present paper is
to provide an {\it analytical} explanation for this ({\it
numerically} observed) characteristic inequality. Moreover, below we
shall derive a stronger lower bound on the dimensionless
charge-to-mass ratio of these explosive (unstable) charged massive
scalar fields.

In order to analyze the physical properties of the composed
RN-black-hole-charged-scalar-field-mirror system, we shall first
express the radial equation (\ref{Eq11}) for the charged massive
scalar fields in the form of a Schr\"odinger-like wave equation. To
this end, it proves useful to define the new radial function
\begin{equation}\label{Eq16}
\psi=rR\  ,
\end{equation}
in terms of which the radial equation (\ref{Eq11}) can be expressed
in the form
\begin{equation}\label{Eq17}
{{d^2\psi}\over{dy^2}}-V\psi=0\  ,
\end{equation}
where the radial coordinate $y$ is defined by the relation
\begin{equation}\label{Eq18}
dy={{dr}\over{f(r)}}\  .
\end{equation}
The effective radial potential in (\ref{Eq17}) is given by
\begin{equation}\label{Eq19}
V=V(r;M,Q,\omega,q,\mu,l)=-\Big(\omega-{{qQ}\over{r}}\Big)^2+{{f(r)H(r)}\over{r^2}}\
,
\end{equation}
where
\begin{equation}\label{Eq20}
H(r;M,Q,\mu,l)= \mu^2r^2+l(l+1)+{{2M}\over{r}}-{{2Q^2}\over{r^2}}\ .
\end{equation}

In the next section we shall analyze the near-horizon properties of
the effective radial potential $V(r)$ that appears in the
Schr\"odinger-like wave equation (\ref{Eq17}) for the charged
massive scalar fields in the charged RN black-hole spacetime. We
shall then use these properties in order to study the near-horizon
spatial behavior of the radial eigenfunction $\psi$ which
characterizes the charged massive scalar fields.

\section{The near-horizon behavior of the charged scalar eigenfunctions}

Our main goal is to explore the onset of superradiant instabilities
in the composed RN-black-hole-charged-scalar-field-mirror system.
Thus, we shall henceforth analyze the behavior of the marginally
stable (stationary) charged field modes (\ref{Eq10}) which mark the
boundary of the superradiant instability regime \cite{Notenww}. In
particular, in this section we shall study the near-horizon spatial
behavior of the radial eigenfunction $\psi$ which characterizes the
stationary (marginally stable) resonances of the charged scalar
fields in the charged RN black-hole spacetime. Specifically, we
shall prove below that this characteristic function is a positive
\cite{Notepp}, increasing, and convex function in the near-horizon
${{r-r_+}\over{r_+-r_-}}\ll1$ region of the RN black-hole spacetime.

To that end, we shall first define the dimensionless variables
\begin{equation}\label{Eq21}
x\equiv {{r-r_+}\over{r_+}}\ \ \ \ ; \ \ \ \
\tau\equiv{{r_+-r_-}\over{r_+}}\  ,
\end{equation}
and study the near-horizon $x\ll\tau$ \cite{Notenex} behavior of the
effective radial potential (\ref{Eq19}). Substituting the
characteristic resonant frequency (\ref{Eq10}) of the marginally
stable charged scalar fields into the expression (\ref{Eq19}) of the
effective radial potential, one finds
\begin{equation}\label{Eq22}
r^2_+V(x\to0)=H(r_+)\tau\cdot x+O[(qQ)^2x^2]\
\end{equation}
in the near-horizon region
\begin{equation}\label{Eq23}
x\ll \tau\times {{H(r_+)}\over{(qQ)^2}}\  ,
\end{equation}
where [see Eq. (\ref{Eq20})]
\begin{equation}\label{Eq24}
H(r=r_+)=\mu^2r^2_++l(l+1)+1-{{Q^2}\over{r^2_+}}\  .
\end{equation}
Remembering that $1-Q^2/r^2_+>0$
, one finds the characteristic inequality
\begin{equation}\label{Eq25}
H(r=r_+)>0
\end{equation}
for the massive charged scalar fields. Equations (\ref{Eq22}) and
(\ref{Eq25}) imply that
\begin{equation}\label{Eq26}
V\geq0
\end{equation}
in the near-horizon region (\ref{Eq23}).

Integrating the relation (\ref{Eq18}) in the near-horizon region,
\begin{equation}\label{Eq27}
x\ll\tau\  ,
\end{equation}
one finds
\begin{equation}\label{Eq28}
y={{r_+}\over{\tau}}\ln(x)+O(x)\  ,
\end{equation}
which implies \cite{Noteyas}
\begin{equation}\label{Eq29}
x=e^{\tau y/r_+}[1+O(e^{\tau y/r_+})].
\end{equation}
Taking cognizance of Eqs. (\ref{Eq17}), (\ref{Eq22}), and
(\ref{Eq29}), one finds the near-horizon $x\ll\tau$ behavior
\begin{equation}\label{Eq30}
{{d^2\psi}\over{d\tilde y^2}}-{{4H(r_+)}\over{\tau}}e^{2\tilde
y}\psi=0\
\end{equation}
of the Schr\"odinger-like wave equation (\ref{Eq17}), where
\begin{equation}\label{Eq31}
\tilde y\equiv {{\tau}\over{2r_+}}y\  .
\end{equation}

The physical solution \cite{Notephs} of the near-horizon
Schr\"odinger-like wave equation (\ref{Eq30}) is given by the
modified Bessel function of the first kind \cite{Abram,Noteab1}:
\begin{equation}\label{Eq32}
\psi(y)=I_0\Big(2\sqrt{{{H(r_+)}\over{\tau}}}e^{\tau y/2r_+}\Big)\
.
\end{equation}
Using the well-known properties of the modified Bessel function
$I_0$ \cite{Abram}, one finds from (\ref{Eq32}) that the radial
eigenfunction $\psi$, which characterizes the charged massive scalar
fields in the charged RN black-hole spacetime, is a positive,
increasing, and convex function in the near-horizon region [see Eqs.
(\ref{Eq23}) and (\ref{Eq27})]
\begin{equation}\label{Eq33}
x\ll\tau\times \min\{1,H(r_+)/(qQ)^2\}\  .
\end{equation}
That is,
\begin{equation}\label{Eq34}
\{\psi>0\ \ \ \text{and}\ \ \ {{d\psi}\over{dy}}>0\ \ \ \text{and}\
\ \ {{d^2\psi}\over{dy^2}}>0\}\ \ \ \ \text{for}\ \ \ \
0<x\ll\tau\times \min\{1,H(r_+)/(qQ)^2\}\ .
\end{equation}

Taking cognizance of the characteristic near-horizon spatial
behavior (\ref{Eq34}) of the radial eigenfunction $\psi$
\cite{Noteep1}, together with the boundary condition (\ref{Eq15})
which is dictated by the presence of the reflecting mirror, one
concludes that the radial eigenfunction $\psi$ must have (at least)
one maximum point, $x=x_{\text{max}}$, between the black-hole
horizon [where $\psi$ is a positive and increasing function, see
(\ref{Eq34})] and the reflecting mirror [where $\psi$ vanishes, see
(\ref{Eq15})]. We note, in particular, that the radial eigenfunction
$\psi$ is characterized by the relations
\begin{equation}\label{Eq35}
\{\psi>0\ \ \ \text{and}\ \ \ {{d^2\psi}\over{dy^2}}<0\}\ \ \
\text{for}\ \ \ x=x_{\text{max}}\
\end{equation}
at the maximum point $x=x_{\text{max}}$.

\section{The superradiant instability regime of the charged black-hole bomb}

It the previous section we have proved that the radial eigenfunction
$\psi$, which characterizes the confined charged scalar fields in
the charged RN black-hole spacetime, must have (at least) one
maximum point, $r=r_{\text{max}}$, between the black-hole horizon
and the reflecting mirror. That is,
\begin{equation}\label{Eq36}
r_+<r_{\text{max}}<r_{\text{m}}\  .
\end{equation}
Taking cognizance of Eqs. (\ref{Eq17}) and (\ref{Eq35}), one finds
that the effective radial potential is characterized by the relation
\begin{equation}\label{Eq37}
V(r=r_{\text{max}})<0\
\end{equation}
at this maximum point. We shall now use this characteristic
inequality in order to derive a generic bound on the superradiant
instability regime of the charged black-hole bomb.

Substituting the characteristic resonant frequency (\ref{Eq10}) of
the marginally stable charged scalar fields into the expression
(\ref{Eq19}) of the effective radial potential, one finds the
relation
\begin{equation}\label{Eq38}
V(r=r_{\text{max}};\omega=\omega_{\text{c}})={{r_{\text{max}}-r_+}\over{r^2_{\text{max}}}}
\Big[{{r_{\text{max}}-r_-}\over{r^2_{\text{max}}}}H(r_{\text{max}})-(qQ)^2{{r_{\text{max}}-r_+}\over{r^2_+}}\Big]\
,
\end{equation}
which yields the inequality [see (\ref{Eq37})]
\begin{equation}\label{Eq39}
{{qQ}\over{r_+}}>\sqrt{{{r_{\text{max}}-r_-}\over{r_{\text{max}}-r_+}}\cdot{{H(r_{\text{max}})}\over{r^2_{\text{max}}}}}\
.
\end{equation}

Using the inequality $r_{\text{max}}<r_{\text{m}}$ [see Eq.
(\ref{Eq36})], one finds
\begin{equation}\label{Eq40}
{{r_{\text{max}}-r_-}\over{r_{\text{max}}-r_+}}>{{r_{\text{m}}-r_-}\over{r_{\text{m}}-r_+}}\
\end{equation}
and
\begin{equation}\label{Eq41}
{{l(l+1)}\over{r^2_{\text{max}}}}>{{l(l+1)}\over{r^2_{\text{m}}}}\
.
\end{equation}
In addition, for charged RN black holes the expression
$2M/r^3-2Q^2/r^4$ is a concave function whose maximum is located at
$r=4Q^2/3M$. One can therefore write \cite{Noteexl}
\begin{equation}\label{Eq42}
{{2M}\over{r^3_{\text{max}}}}-{{2Q^2}\over{r^4_{\text{max}}}}> {\cal
F}\equiv
\begin{cases}
{{r_+-r_-}\over{r^3_+}} &\ \text{for}\ \ \ \ r_{\text{m}}\leq4Q^2/3M\ ; \\
\min\{{{r_+-r_-}\over{r^3_+}},{{2M}\over{r^3_{\text{m}}}}-{{2Q^2}\over{r^4_{\text{m}}}}\}
&\ \text{for}\ \ \ \ r_{\text{m}}>4Q^2/3M\ .
\end{cases}
\end{equation}
From Eqs. (\ref{Eq20}), (\ref{Eq41}), and (\ref{Eq42}), one finds
the lower bound
\begin{equation}\label{Eq43}
{{H(r_{\text{max}})}\over{r^2_{\text{max}}}}>
\mu^2+{{l(l+1)}\over{r^2_{\text{m}}}}+{\cal F}\  .
\end{equation}

Substituting the inequalities (\ref{Eq40}) and (\ref{Eq43}) into
(\ref{Eq39}), one can write the lower bound on the dimensionless
quantity $qQ$ in terms of the physical parameters
$\{r_{\pm},r_{\text{m}}\}$ of the black hole and its confining
mirror:
\begin{equation}\label{Eq44}
qQ>\sqrt{{{r_{\text{m}}-r_-}\over{r_{\text{m}}-r_+}}\cdot\Big[\mu^2+{{l(l+1)}\over{r^2_{\text{m}}}}+{\cal
F}\Big]r^2_+}\  .
\end{equation}
It is worth emphasizing that the analytically derived lower bound
(\ref{Eq44}) provides a necessary condition for the development of
the superradiant instabilities in the composed
RN-black-hole-charged-scalar-field-mirror system
\cite{Notecplx,Notecplx2}.

\section{Numerical confirmation}

We shall now verify the validity of the analytically derived lower
bound (\ref{Eq44}) on the superradiant instability regime of the
charged black-hole bomb. The instability spectrum of this composed
RN-black-hole-charged-massive-scalar-field-mirror system was
investigated numerically in \cite{Dego}. In Table \ref{Table1} we
display the dimensionless ratio
$(qQ)^{\text{stat}}/(qQ)^{\text{bound}}$, where $(qQ)^{\text{stat}}$
is the {\it numerically} computed \cite{Dego} value of the quantity
$qQ$ which corresponds to the stationary (marginally stable) charged
scalar configurations \cite{Notesbo}, and $(qQ)^{\text{bound}}$ is
the {\it analytically} derived lower bound on the superradiant
instability regime given by Eq. (\ref{Eq44}). One finds from Table
\ref{Table1} that the charged black-hole bomb is characterized by
the relation $(qQ)^{\text{stat}}/(qQ)^{\text{bound}}>1$, in
agreement with the analytically derived lower bound (\ref{Eq44}).

\begin{table}[htbp]
\centering
\begin{tabular}{|c|c|c|c|}
\hline $Q/M$ & \ \ 0.990 \ \ & \ \ 0.997 \ \ & \ \ 0.999 \ \ \ \\
\hline \ \ $(qQ)^{\text{stat}}/(qQ)^{\text{bound}}$
\ \ &\ \ \ 1.03\ \ \ \ &\ \ \ 1.09\ \ \ \ &\ \ \ 1.12\ \ \ \ \\
\hline
\end{tabular}
\caption{The superradiant instability regime of the composed
RN-black-hole-charged-scalar-field-mirror system (the charged
black-hole bomb). We display the dimensionless ratio
$(qQ)^{\text{stat}}/(qQ)^{\text{bound}}$, where $(qQ)^{\text{stat}}$
is the {\it numerically} computed \cite{Dego} value of the quantity
$qQ$ which corresponds to the stationary (marginally stable) charged
scalar configurations \cite{Notesbo}, and $(qQ)^{\text{bound}}$ is
the {\it analytically} derived lower bound on the superradiant
instability regime given by Eq. (\ref{Eq44}). The data presented is
for the case $M\mu=0.3, Mq=0.36$, and $l=1$. One finds that the
superradiant instability regime of the charged black-hole bomb is
characterized by the relation
$(qQ)^{\text{stat}}/(qQ)^{\text{bound}}>1$, in agreement with the
analytically derived lower bound (\ref{Eq44}).} \label{Table1}
\end{table}

\section{Summary and discussion}

We have studied analytically the superradiant instability regime of
the charged black-hole bomb. This physical system is composed of a
charged massive scalar field which, on the one hand, extracts the
Coulomb energy of a charged Reissner-Nordstr\"om black hole and, on
the other hand, is prevented from escaping to infinity by a
reflecting mirror which surrounds the black hole. We have proved
that in order for the superradiant instability to develop in this
composed charged black-hole bomb, the dimensionless quantity $qQ$ of
the black-hole-field system must be bounded from below as in
(\ref{Eq44}).

In a very interesting study, Degollado et. al. \cite{Dego} have used
numerical techniques to study the instability spectrum of the
charged black-hole bomb. In particular, it was found in \cite{Dego}
that the inequality $q/\mu>1$ [see Eq. (\ref{Eq4})] provides a
necessary condition for the development of the superradiant
instability in this composed system. We can now provide an {\it
analytical} explanation for this {\it numerically} observed
\cite{Dego} necessary condition: Using the relation $Q^2=r_+r_-$,
one finds from (\ref{Eq44}) the compact lower bound \cite{Notehw}
\begin{equation}\label{Eq45}
{{q}\over{\mu}}>\sqrt{{{r_{\text{m}}/r_--1}\over{r_{\text{m}}/r_+-1}}}>1
\end{equation}
on the dimensionless charge-to-mass ratio of the scalar fields in
the explosive (unstable) regime of the charged black-hole bomb
\cite{Notelrm,Notermm}. It is worth emphasizing that this lower
bound provides a necessary condition for the development of the
superradiant instabilities in the composed
RN-black-hole-charged-scalar-field-mirror system \cite{Notesw}.

Thus far, we have treated the composed charged black-hole bomb at
the {\it classical} level. It should be emphasized, however, that
the well known Schwinger {\it quantum} pair-production mechanism
\cite{Schw1,Schw2,Schw3,Schw4} restricts the physical parameters of
the composed RN-black-hole-charged-massive-scalar-field system. In
particular, this vacuum polarization effect sets the upper bound
\cite{Schw1,Schw2,Schw3,Schw4,NoteSch} $E_+\ll E_{\text{c}}\equiv
\mu^2/q\hbar$ on the strength of the black-hole electric field (here
$E_+=Q/r^2_+$ is the electric field at the horizon of the charged RN
black hole). The quantum production of charged
particle/anti-particle pairs in the charged black-hole spacetime
(the Schwinger discharge of the RN black hole) therefore sets the
upper bound $qQ\ll\mu^2r^2_+$ on the physical parameters of the
composed black-hole-field system. Taking cognizance of Eq.
(\ref{Eq44}) \cite{Noterh} one finds that, in the superradiant
explosive regime, the dimensionless quantity $qQ$ is restricted by
the two inequalities
\begin{equation}\label{Eq46}
\mu r_+<qQ\ll\mu^2r^2_+\  .
\end{equation}

The two inequalities in (\ref{Eq46}) imply that, in physically
acceptable situations \cite{Notets}, the explosive charged massive
scalar fields must be characterized by the strong inequalities
\begin{equation}\label{Eq47}
1\ll\mu r_+<qQ\  .
\end{equation}
It is worth noting that the physical restriction $\mu r_+\gg1$ [see
(\ref{Eq47})] imposed by the quantum Schwinger pair-production
mechanism (the vacuum polarization effect) implies that, in
physically acceptable situations \cite{Notets}, the lower bound
(\ref{Eq44}) is well approximated by the lower bound (\ref{Eq45})
\cite{Notesw2}.

Finally, it is worth emphasizing again that in this study we have
treated the charged massive scalar fields at the linear level. Our
analytical results are therefore expected to be valid in the early
stages of the development of the superradiant instability (that is,
in the {\it ignition} stage of the black-hole bomb), when the
external charged scalar fields are weak and can still be regarded as
perturbation fields on the background of the charged RN black-hole
spacetime. As we demonstrated explicitly in this paper, the main
advantage of this perturbative (linearized) approach stems from the
fact that the physical properties of the composed
RN-black-hole-charged-massive-scalar-field-mirror system (the
charged black-hole bomb) can be explored {\it analytically} in the
linear regime. It should be emphasized, however, that the late-time
(non-linear) development of the superradiant instability can only be
tackled with {\it numerical} techniques, as recently done in the
interesting numerical work of Sanchis-Gual et. al. \cite{Herrec}.


\bigskip
\noindent
{\bf ACKNOWLEDGMENTS}
\bigskip

This research is supported by the Carmel Science Foundation. I thank
Yael Oren, Arbel M. Ongo, Ayelet B. Lata, and Alona B. Tea for
stimulating discussions.



\begin{thebibliography}{99}
\bibitem{Noteun} We use natural units in which $G=c=\hbar=1$.
\bibitem{Notelin} That is, in the regime of test (linearized) charged scalar
fields propagating on the background of the charged RN black-hole
spacetime.
\bibitem{Noterr} Here $r$ is the Schwarzschild areal coordinate.
\bibitem{Noteuni} It is worth noting that the physical parameters $\mu$ and $q$ of the charged massive scalar field
stand respectively for $\mu/\hbar$ and $q/\hbar$. Hence, these field
parameters have the dimensions of $($length$)^{-1}$.
\bibitem{Noteqp} We shall assume, without loss of generality, that
$q>0$ and $Q>0$.
\bibitem{Notelm} Here $l$ and $m$ are respectively the spherical harmonic index and
the azimuthal harmonic index of the field mode.
\bibitem{Noteom} We shall henceforth omit the angular harmonic indices $\{l,m\}$ for brevity.
\bibitem{Notenww} Note that these stationary field modes, which are characterized by the critical
(marginal) frequency (\ref{Eq10}) for the superradiant scattering
phenomenon, mark the boundary between stable ($\Im\omega<0$) and
unstable ($\Im\omega>0$) resonances of the confined scalar fields.
\bibitem{Notepp} One can assume $R(r=r_+)\geq0$ without loss of generality.
\bibitem{Notenex} We shall henceforth consider non-extremal black
holes which are characterized by $\tau>0$.
\bibitem{Noteyas} It is worth noting that the near-horizon $x\ll\tau$ region corresponds to $\tau y/r_+\to
-\infty$ [see Eq. (\ref{Eq28})]. This implies $e^{\tau y/r_+}\to 0$
in the near-horizon region.
\bibitem{Notephs} That is, the radial solution which respects the physically motivated boundary
condition (\ref{Eq14}).
\bibitem{Noteab1} Here we have used equations (9.1.54) and (9.6.3) of \cite{Abram}.
\bibitem{Noteep1} It is worth emphasizing again that the inequalities presented in (\ref{Eq34})
imply that the radial eigenfunction $\psi$ is a positive and
increasing function in the vicinity of the black-hole horizon [that
is, in the near-horizon $x\ll\tau\times \min\{1,H(r_+)/(qQ)^2\}$
region].
\bibitem{Noteexl} Here we have used the fact that $2M/r^3-2Q^2/r^4$ is a monotonic increasing function
in the regime $r<4Q^2/3M$. Thus,
$2M/r^3_{\text{max}}-2Q^2/r^4_{\text{max}}>2M/r^3_+-2Q^2/r^4_+=(r_+-r_-)/r^3_+$
in the regime $r_+<r_{\text{max}}<r_{\text{m}}\leq 4Q^2/3M$. On the
other hand, the expression $2M/r^3-2Q^2/r^4$ is a monotonic
decreasing function in the regime $r>4Q^2/3M$. Thus, for
$r_{\text{m}}>4Q^2/3M$ one cannot conclude that $r_{\text{max}}$ is
located in the increasing region of $2M/r^3-2Q^2/r^4$, in which case
one can only write $2M/r^3_{\text{max}}-2Q^2/r^4_{\text{max}}>\min
\{{{2M}/{r^3_+}}-{{2Q^2}/{r^4_+}},
{{2M}/{r^3_{\text{m}}}}-{{2Q^2}/{r^4_{\text{m}}}}\}$.
\bibitem{Notecplx2} It is worth emphasizing that we have proved in this paper that
the lower bound (\ref{Eq44}) on the dimensionless quantity $qQ$
characterizes the marginally stable (stationary) field modes
(\ref{Eq10}). In addition, we know that the RN-mirror system is {\it
stable} to uncharged field modes with $qQ=0<(qQ)^{\text{bound}}$,
see Eq. (\ref{Eq3}) [Here $(qQ)^{\text{bound}}$ denotes the r.h.s of
(\ref{Eq44})]. It is therefore natural to expect (though it should
be emphasized that we have not proved it here) that the {\it
unstable} charged field modes are characterized by the opposite
inequality $qQ>(qQ)^{\text{bound}}$.
\bibitem{Notesbo} These marginally stable (stationary) field configurations
[with $\omega=\omega_{\text{c}}$, see Eq. (\ref{Eq10})] mark the
onset of superradiant instabilities in the composed
RN-black-hole-charged-massive-scalar-field-mirror system. In
particular, the explosive (unstable) regime of the charged
black-hole bomb is characterized by confined charged field
configurations whose dimensionless quantity $qQ$ lies in the regime
$qQ>(qQ)^{\text{stat}}$ (see also \cite{Dego}).
\bibitem{Notehw} Here we have used the inequalities $l(l+1)\geq 0$
and ${\cal F}\geq0$ [see Eq. (\ref{Eq42})] in Eq. (\ref{Eq44}).
\bibitem{Notelrm} Note that the analytically derived lower bound (\ref{Eq45}) on the
dimensionless charge-to-mass ratio of the explosive scalar fields
can be approximated by the simple inequality $q/\mu>1$ in the regime
$r_{\text{m}}/r_+\gg1$ of large mirror radii.
\bibitem{Notermm} Note that the lower bound (\ref{Eq45}) on the dimensionless charge-to-mass
ratio of the explosive scalar fields can also be expressed as a
lower bound on the dimensionless radius of the confining mirror:
${{r_{\text{m}}}\over{r_+}}>{{(q/\mu)^2-1}\over{(q/\mu)^2-r_+/r_-}}$.
\bibitem{Notesw} It is worth emphasizing that the lower bound
(\ref{Eq44}) is stronger than the lower bound (\ref{Eq45}).
\bibitem{NoteSch} It is worth noting that explicit calculations
made in \cite{Schw4} reveal that the quantum Schwinger pair
production mechanism of the charged massive particles becomes
effective already at $E\simeq 0.03E_{\text{c}}\ll E_{\text{c}}$.
\bibitem{Noterh} Note, in particular, that the r.h.s of (\ref{Eq44}) is larger than
$\mu r_+$, which implies $qQ>\mu r_+$ in the superradiant explosive
regime of the charged black-hole bomb.
\bibitem{Notets} That is, taking into account the well known quantum Schwinger pair-production
mechanism in the charged black-hole spacetime
\cite{Schw1,Schw2,Schw3,Schw4}.
\bibitem{Notesw2} It is worth emphasizing again that the lower bound
(\ref{Eq44}) is stronger than the lower bound (\ref{Eq45}).
\end{thebibliography}
\end{document}